\documentclass[11pt,a4paper]{article}
\usepackage{jstyle}

\usepackage[utf8]{inputenc}
\usepackage{amsthm,amsmath,latexsym,amssymb,amsfonts,amssymb,amscd}
\usepackage{hyperref}
\usepackage{color}
\usepackage{mathabx}
\usepackage[compat=1.1.0]{tikz-feynman}
\usepackage[english]{babel}

\definecolor{rougef}{rgb}{0.56,0,0}
\definecolor{vertf}{rgb}{0,0.5,0}
\definecolor{bleuf}{rgb}{0,0,0.8}
\definecolor{violetf}{rgb}{0.5,0,0.5}

\newcommand{\vac}{\Omega}%\newcommand{\vac}{|0\rangle}
\newcommand{\vacl}{\langle \Omega,}%\newcommand{\vacl}{\langle 0|} 
\newcommand{\vacr}{\Omega \rangle}%\newcommand{\vacr}{|0\rangle}

\usepackage[colorinlistoftodos,draft]{todonotes} 
\author[a,b]{Stefan Fredenhagen,}
\author[a]{Olaf Kr\"uger}
\author[c]{and Karapet Mkrtchyan}

\affiliation[a]{University of Vienna, Faculty of Physics,\\ Boltzmanngasse 5, 1090 Vienna, Austria}
\affiliation[b]{Erwin Schr{\"o}dinger International Institute for Mathematics and Physics,\\ University of Vienna, Boltzmanngasse 9, 1090 Vienna, Austria}
\affiliation[c]{Max Planck Institute for Gravitational Physics (Albert Einstein Institute)\\
Am M\"uhlenberg 1, 14476 Potsdam, Germany}

\emailAdd{stefan.fredenhagen@univie.ac.at}
\emailAdd{olaf.krueger@univie.ac.at}
\emailAdd{karapet.mkrtchyan@aei.mpg.de}
\title{\centering
\LARGE{Constraints for Three-Dimensional Higher-Spin Interactions and Conformal Correlators}
}

\abstract{In the context of higher-spin holography, we compare the classification of cubic interaction vertices for higher-spin gravity theories in three dimensions to the possible three-point correlation functions of conserved higher-spin currents in two-dimensional conformal field theories. In both cases, the allowed structures are governed by triangle inequalities for the involved spins. It is established that higher-order correlators satisfy similar polygon inequalities and that the same inequalities are valid for higher-order continuations of cubic vertices in the three-dimensional higher-spin gravity.}

\begin{document}

\maketitle

%%%%%%%%%%%%%%%%%%%%%%%%%%%%%%%%%%%%%%%%%%%%%%%%%%%%%%%%%%%%%
%%%%%%%%%%%%%%%%%%%%%%%%%%%%%%%%%%%%%%%%%%%%%%%%%%%%%%%%%%%%%
\section{Introduction}
%%%%%%%%%%%%%%%%%%%%%%%%%%%%%%%%%%%%%%%%%%%%%%%%%%%%%%%%%%%%%
%%%%%%%%%%%%%%%%%%%%%%%%%%%%%%%%%%%%%%%%%%%%%%%%%%%%%%%%%%%%%

Higher-spin (HS) gravity theories \cite{Vasiliev:1990en,Prokushkin:1998bq,Vasiliev:2003ev} (for reviews see e.g.\ \cite{bciv,Campoleoni:2011tn,Didenko:2014dwa,Rahman:2015pzl,Kessel:2017mxa}) provide interesting extensions of gravity for which the holographic duality can in principle be studied perturbatively in contrast to the string theoretic AdS/CFT duality. They are therefore a promising class of gravitational theories for which one can hope to understand the AdS/CFT correspondence on the quantum level and to learn about quantum gravity from the quantum field theory on the boundary. Higher-spin holography (for a review see~\cite{Giombi:2016ejx}) is to a large extent controlled by an enormous symmetry: the higher-spin gauge symmetry in the bulk and the corresponding higher-spin global symmetry on the boundary \cite{Joung:2014qya}. This symmetry becomes particularly powerful when the boundary theory is two-dimensional and the asymptotic symmetry algebra of a higher-spin gauge theory on asymptotically AdS$_{3}$ space-times becomes an infinite-dimensional $\mathcal{W}$-algebra~\cite{Henneaux:2010xg,Campoleoni:2010zq} (see also \cite{Gaberdiel:2011wb,Campoleoni:2011hg,Campoleoni:2014tfa,Joung:2017hsi}). Studying conformal field theories (CFTs) with such $\mathcal{W}$-algebras as chiral symmetry algebras has led to a concrete proposal for a holographic duality between Prokushkin-Vasiliev higher-spin gauge theory~\cite{Prokushkin:1998bq} and limits of minimal model CFTs~\cite{Gaberdiel:2010pz} (for a review see~\cite{Gaberdiel:2012uj}).

In all concrete proposals for higher-spin dualities one of the problems encountered is that there is no standard action known for the higher-spin bulk theory, which makes quantisation and thus a quantum analysis of holography difficult. An attempt to construct such an action perturbatively is the Noether-Fronsdal program which starts out from the free, quadratic Fronsdal action \cite{Fronsdal:1978rb} of higher-spin gauge fields and adds interactions order by order while preserving gauge invariance. Successful implementation of this program may provide a classification of all possible higher-spin gauge theories which admit a perturbative formulation.\footnote{The current status of the Noether-Fronsdal program in arbitrary dimensions is reviewed in \cite{Mkrtchyan:2017ixk,Kessel:2018ugi}.}
In three dimensions, a Chern-Simons formulation \cite{Blencowe:1988gj} is available for HS Gravity without matter (see also \cite{Campoleoni:2010zq,Gwak:2015vfb,Gwak:2015jdo}), allowing an alternative derivation of the action in metric variables. This approach was taken in  \cite{Campoleoni:2012hp,Fredenhagen:2014oua} to directly obtain an action in  terms of Fronsdal fields for Chern-Simons spin-three theories \cite{Campoleoni:2010zq}.

In three dimensions, the classification of gauge invariant cubic interaction vertices has recently been completed~\cite{Mkrtchyan:2017ixk,Kessel:2018ugi}. 
 The purpose of this paper is to establish that this classification matches the structures of three-point functions of conserved currents in a two-dimensional CFT. In that sense, rather than checking the AdS/CFT correspondence in specific examples, we study the structure of general higher-spin theories and general CFTs. We thus extend the known match between cubic bulk vertices and boundary correlators in (boundary) dimensions three and higher~\cite{Metsaev:2005ar,Manvelyan:2010jr,Giombi:2011rz,Costa:2011mg,Joung:2011ww,Conde:2016izb,Francia:2016weg,Kravchuk:2016qvl} to two dimensions. 

Before we summarise our main results, we briefly want to explain why the analysis in two dimensions is different and not just more or less a corollary of the higher-dimensional case. On the bulk side, the analysis is complicated by the appearance of many dimension-dependent identities (Schouten identities) that are special to three (bulk) dimensions. On the boundary, a particular feature of two-dimensional CFTs is that the conserved currents are chiral or anti-chiral fields. In higher dimensions, the product of Lorentz covariant conserved traceless currents
\begin{equation}
\hat J_{\m_1\dots\m_n} (x) =J_{(\m_1\dots\m_k}(x)\tilde{J}_{\m_{k+1}\dots\m_{n})}(x)-(\text{traces})
\end{equation}
does not define a conserved current, but in two dimensions the product of two chiral currents is again chiral and hence conserved. Therefore we have a large freedom of non-linear redefinitions of the basis fields which can change the correlators. Contrary to higher dimensions we therefore do not just have to analyse the possible structure of correlation functions from conformal invariance (which is trivial in two dimensions), but we have to analyse what the independent structures are modulo non-linear changes of the basis. 
\smallskip

The analysis of~\cite{Mkrtchyan:2017ixk,Kessel:2018ugi} has shown that for three-dimensional higher-spin gauge fields with spins $s_{1},s_{2},s_{3}\geq 2$, there is precisely one parity-even and one parity-odd cubic vertex whenever the spins satisfy triangle inequalities
\begin{equation}
|s_{1}-s_{2}|< s_{3} < s_{1}+s_{2}\, ,
\end{equation}
and none otherwise. The main results of this paper are
\begin{enumerate}
\item In any $\mathcal{W}$-algebra, there is a unique basis of quasi-primary fields for which the $r$-point correlators vanish unless the involved spins $s_{1},\dotsc ,s_{r}$ satisfy polygon inequalities\footnote{The name of these inequalities comes from a geometric interpretation: they are equivalent to a statement that $r$ line segments with lengths $s_i-1$ can form a polygon on a plane.}
\begin{equation}
s_{i} < \sum_{j=1,j\not =i}^{r} s_{j} -r +3 \, .
\end{equation}  
In particular, in this basis the three-point functions obey the same triangle inequalities as the higher-spin cubic vertices in three dimensions.
\item All higher-order bulk vertices that depend on the cubic vertices, in the sense that they are needed to complete the cubic vertices to a consistent theory, obey the same polygon inequalities.\footnote{Examples of higher-order couplings that are completions of cubic ones are the quartic vertex of Yang-Mills theory and all higher-order vertices in Einstein Gravity: they are needed for the consistency of the cubic deformation, but do not introduce new parameters and necessarily vanish when the coefficient of the cubic coupling goes to zero.}
\end{enumerate}
We thus find a perfect match of structures in the bulk and the boundary at this level. Note that at this point we have not implemented the consistency of the gauge algebra in the bulk or the associativity of operator products on the boundary. These will give additional restrictions on both sides of the duality.
\smallskip

The outline of the paper is as follows. We start in Section~\ref{NFRev} by a review of the classification of cubic higher-spin vertices in three dimensions. We then give a short review on the structure of correlation functions in two-dimensional CFTs in Section~\ref{CFTRev}. In the subsequent Section~\ref{sec:CFTtriangle} we then describe in detail how one can construct a basis for which the correlation functions vanish unless the involved spins satisfy polygon inequalities. This basis necessarily consists of primary fields (except possibly a quasi-primary spin-$2$ field), and it is uniquely fixed by demanding that the polygon inequalities are obeyed. Motivated by the CFT results, we then show in Section~\ref{sec:HSpolygon} that all dependent higher-order bulk vertices satisfy the same polygon inequalities. We end with a brief discussion.

%%%%%%%%%%%%%%%%%%%%%%%%%%%%%%%%%%%%%%%%%%%%%%%%%%%%%%%%%%%%%
%%%%%%%%%%%%%%%%%%%%%%%%%%%%%%%%%%%%%%%%%%%%%%%%%%%%%%%%%%%%%
\section{Cubic Couplings in Three Dimensions: Review}\label{NFRev}
%%%%%%%%%%%%%%%%%%%%%%%%%%%%%%%%%%%%%%%%%%%%%%%%%%%%%%%%%%%%%
%%%%%%%%%%%%%%%%%%%%%%%%%%%%%%%%%%%%%%%%%%%%%%%%%%%%%%%%%%%%%

The Noether-Fronsdal program is a systematic procedure to construct possible covariant actions for higher-spin gravity theories order by order. In this section, we review the recent results \cite{Mkrtchyan:2017ixk,Kessel:2018ugi} on cubic couplings in a three-dimensional theory.

The starting point is the free Lagrangian given as a sum of quadratic Fronsdal Lagrangians:
\begin{equation}
\mathcal{L}^{(2)}=\sum_{k=1}^{N}\mathcal{L}^{(2)}_{s_k}[\vf^{(k)}]\,,
\end{equation}
where $\mathcal{L}^{(2)}_{s}$ is the spin-$s$ Fronsdal Lagrangian, $k$ is a label counting the fields $\vf^{(k)}$, and $s_k$ is the spin of the $k$-th field. The number $N$ of all fields involved can be arbitrary, or even infinite as it is in the case of Prokushkin-Vasiliev theory --- a theory of HS Gravity coupled to matter \cite{Prokushkin:1998bq}.

For vector fields, we have two different possibilities for the free action: The Maxwell action, which is a
particular case of the Fronsdal action for $s=1$, and a (abelian) Chern-Simons action of the form
$\int A\wedge dA$. The former is equivalent to a scalar, while the latter carries no bulk degrees of freedom. In a
general theory, there may be both types of fields in the spectrum.

A spin $s\geq 2$ field is described by a rank $s$ symmetric Lorentz tensor $\vf_{\m_1\dots\m_s}$. The explicit form of the Fronsdal action can be found in \cite{Fronsdal:1978rb}.
We will sketch the idea of the construction here  and briefly discuss what is known so far.

The main assumption is that the theory is given by a local Lagrangian,\footnote{This assumption may be too strong
  and leave out interesting theories with HS spectrum. In particular, it is not clear if Prokushkin-Vasiliev
  theory \cite{Prokushkin:1998bq} is ruled out by this assumption. It is sufficient for our purposes here to
  restrict ourselves to strictly local Lagrangian functionals since our conclusions concern the gauge sector of
  three dimensional HS Gravity theories excluding scalar and Maxwell fields.} non-linear in the fields
$\vf^{(k)}$, which can be constructed order by order in a small parameter $g$. The latter is usually associated
with an overall factor in front of the cubic Lagrangian:
\begin{align}\label{lagrangian}
\mathcal{L}=\mathcal{L}^{(2)}+g\,\mathcal{L}^{(3)}+g^2\mathcal{L}^{(4)}+\dots\,
\end{align}

This Lagrangian should be gauge invariant with respect to gauge transformations of all fields up to total derivative terms. The gauge transformations can be expanded in powers of the fields:
\begin{align}\label{gaugetransf}
\d \vf^{(k)}=\d^0 \vf^{(k)}+g\, \d^1 \vf^{(k)}+g^2\,\d^2 \vf^{(k)}+\dots\,,
\end{align}
where $\d^n\vf^{(k)}$ is given as a sum of different terms, linear in one of the gauge parameters and of order $n$ in the number of fields. The zeroth order gauge transformation is that of a free field given by (we use symmetrisation with weight one)
\begin{align}
\d^0 \vf^{(k)}_{\m_1\dots\m_{s_k}}=s_k\,\nabla_{(\m_1}\e^{(k)}{}_{\!\!\!\!\!\!\!\m_2\dots\m_{s_k})}\,.
\end{align}
In the following, we will use the shorthand notation $\e^{(k)}$ for the parameter associated to the field $\vf^{(k)}$ with spin $s_k\geq 1$. The higher-order transformation $\delta^{n}\vf^{(k)}$ $(n>0)$ of a field of spin $s_{k}$ can contain gauge parameters of all the other fields.   
The simplest  examples of such theories are Yang-Mills theory and Gravity with or without matter. There, the full structure of higher-order interactions is fixed by the cubic ones.
We assume this is true also for the example under consideration: in the space of all possible non-linear theories, we will restrict ourselves to those for which the full interaction structure is fixed uniquely by the cubic order interaction.
We will study independent higher-order interactions elsewhere.

The higher-order consistency of the gauge symmetry severely constrains the cubic couplings. We assume that there will remain at least one free parameter.
In particular, since we are discussing a theory of Gravity, we can assume that the coefficient of the
Einstein-Hilbert cubic vertex is non-zero. Hence, we can take its coefficient, the Newton constant, as the
parameter $g$ of the expansion, while rescaling all the other coupling constants for cubic vertices by $g_{k,k',k''}\to g\, g_{k,k',k''}$ to have an overall factor $g$ in front of the cubic Lagrangian.

The $n$-th order Lagrangian can be expanded in the basis of vertices of interactions between fields $\vf^{(k_1)}\,,\dots,\, \vf^{(k_n)}$:
\begin{align}
\mathcal{L}^{(n)}=\sum_{k_i,m} g^m_{k_1,\dots,k_n}\,V^m_{k_1,\dots,k_n}\left(\vf^{(k_1)}\,,\dots,\, \vf^{(k_n)}\right)\,,
\end{align}
where $m$ counts the independent coefficients. 
Here, the vertices are by convention symmetric in the field labels, i.e.\ an $n$-point vertex for $n$ pairwise different fields appears $n!$ times in the sum.

Gauge invariance of the theory given by the Lagrangian~\eqref{lagrangian} under the
transformation~\eqref{gaugetransf} has to hold order by order. In particular, at cubic order in the fields (first order in the expansion parameter) we get the cubic Noether equations 
\begin{align}
\d^1 \mathcal{L}^{(2)}+\d^0 \mathcal{L}^{(3)} = 0 \ \text{up to total derivatives} \quad \Longrightarrow  \quad  \d^0 \mathcal{L}^{(3)}\approx 0\,.\label{cubicNE}
\end{align}
Here, the last equality is up to terms proportional to the free equations of motion and up to total derivatives. This can be understood by considering the contribution of the variation of the free Lagrangian, which is schematically
\begin{align}
\d^1 \mathcal{L}^{(2)}=\sum_k \d^1\varphi^{(k)}\mathcal{F}(\varphi^{(k)}) \approx 0 \,,
\end{align}
where $\mathcal{F}(\varphi^{(k)})$ is the free equation of motion for the field $\varphi^{(k)}$. 

In the Noether procedure, we often have to consider equivalence classes with respect to $\approx$, and they enter the analysis of cubic vertices in two distinct ways. First, they appear when we consider gauge invariance as we have just seen. Second, one can add to the cubic vertex any total derivative without changing the action, and any local terms
 proportional to the free field equations by appropriate quadratic field redefinitions: $\varphi\rightarrow
 \varphi+g\,O(\varphi^2)$. Given that the free field equations satisfy the lowest order gauge invariance, $\d^0
 \mathcal{F}(\varphi)=0$, the cubic terms induced by these field redefinitions automatically solve the
 equation~\eqref{cubicNE}. Fixing this redundancy is essential for parametrising the independent non-trivial
 cubic deformations. 

Inversely, fixing the freedom to add terms proportional to the free equations of motion to the cubic vertex also
fixes the freedom of field redefinitions at quadratic order in the fields. Higher-order field redefinitions
are however still allowed: at each order $n\geq 2$ in the field variables, $O(\varphi^n)$, we have the freedom to
redefine $\varphi\rightarrow \varphi + g^{n-1} O(\varphi^n)$ and, correspondingly, the freedom to add terms to the $(n+1)$st order Lagrangian $\mathcal{L}^{(n+1)}$ that are proportional to the free field equations.

At cubic order, there is a natural way to fix the freedom of field redefinitions in a Lorentz covariant manner by ruling out all the terms in the traceless-transverse part of the cubic Lagrangian, where derivatives are contracted to each other. Details can be found in~\cite{Mkrtchyan:2017ixk}.

Fixing this field redefinition freedom for any triple of fields, we are left with two independent deformations --- a parity-even vertex and a parity-odd one --- if the spins of the fields satisfy the triangle inequalities 
\begin{align}
|s_{k_2}-s_{k_3}|<s_{k_1}<s_{k_2}+s_{k_3}\,,\label{TI}
\end{align}
and none otherwise~\cite{Mkrtchyan:2017ixk,Kessel:2018ugi}. Exceptions are the cases with scalar and Maxwell fields, where one has current type couplings.  

Having found the cubic vertices, one can then go back to the Noether equations~\eqref{cubicNE}  for the gauge symmetry at cubic order,
\begin{align}
\d^1 \mathcal{L}^{(2)}+\d^0 \mathcal{L}^{(3)}= \text{total derivative}\,.
\end{align}
Explicitly they can be written as
\begin{equation}\label{cancellation}
\sum_{k, k',k''} \Big( \d^1_{k, k'} \vf^{(k'')} \mathcal{F}(\vf^{(k'')})+ 3\sum_{m}g^{m}_{k,k',k''}\,V^{m}_{k,k',k''} (\d^0\vf^{(k)},\vf^{(k')},\vf^{(k'')})\Big)= \text{total derivative}\,,
\end{equation}
where $\d^1_{k,k'}\vf^{(k'')}$ is the gauge variation of the field $\vf^{(k'')}$, with a parameter $\e^{(k)}$
and linear in the field $\vf^{(k')}$ (we do not need the explicit form here). The contribution containing a specific gauge parameter $\e^{(k)}$ and two fields with fixed labels $k'$ and $k''$ has to vanish separately, therefore we have
\begin{multline}
\d^1_{k, k'} \vf^{(k'')} \mathcal{F}(\vf^{(k'')})+\d^1_{k, k''}\vf^{(k')} \mathcal{F}(\vf^{(k')})+6\sum_m g^m_{k,k',k''}V^m_{k,k',k''}(\d^0\vf^{(k)},\vf^{(k')},\vf^{(k'')})\\
=\text{total derivative}\,.\label{Del1eq}
\end{multline}
The contribution of the cubic vertex in~\eqref{Del1eq} can be compensated by a deformation of the gauge transformation of the form
\begin{align}
\d^1_{k,k'}\vf^{(k'')}=\sum_m\, g^m_{k,k',k''}\; d^m_{k,k',k''}(\e^{(k)},\vf^{(k')})\,,\label{def1}
\end{align}
where $d^m_{k,k',k''}$ is a differential operator acting on its arguments. Eq.~\eqref{Del1eq} shows in particular that the deformations $\d^1_{k,k'}\vf^{(k'')}$ of the gauge transformations satisfy the same triangle inequalities~\eqref{TI} as the cubic vertices: they can be non-zero only when
\begin{equation}
|s_{k}-s_{k'}| < s_{k''} < s_{k} + s_{k'} \, .
\end{equation}
As explained in \cite{Joung:2014qya}, the first order gauge transformations $\d^1_{k,k'}\vf^{(k'')}$ form the algebra of global symmetries of the theory for Killing parameters satisfying $\nabla_{(\m_1}\e^{(k)}{}_{\m_2\dots\m_{s_k})}=0$.
\smallskip

To summarise, after fixing the freedom of field redefinitions at cubic level as in
\cite{Manvelyan:2010wp,Manvelyan:2010jr}, there are two vertices --- parity even \cite{Mkrtchyan:2017ixk} and
parity odd \cite{Kessel:2018ugi} --- for each triple of fields precisely if the spins of the fields satisfy triangle inequalities~\eqref{TI}.\footnote{This is in striking difference to the situation in higher dimensions, where violations of triangle inequalities are possible, in particular, through Bell-Robinson type current couplings \cite{Manvelyan:2009vy}. Such currents do not define non-trivial couplings in three dimensions due to on-shell triviality of the de Wit-Freedman curvatures of Fronsdal fields in $d=3$.}
The precise form of the cubic vertices is not important for the purpose of this article, explicit expressions can be found in~\cite{Mkrtchyan:2017ixk,Kessel:2018ugi}.  
Each of these vertices induces corresponding deformations of the gauge transformations \eqref{def1}. Since the transformations \eqref{def1} stem from cubic vertices, they also vanish for spins $s_k, s_{k'}, s_{k''}$ that violate the triangle inequalities.
In Section~\ref{sec:HSpolygon}, we will use this to prove certain properties of higher-order vertices.

%%%%%%%%%%%%%%%%%%%%%%%%%%%%%%%%%%%%%%%%%%%%%%%%%%%%%%%%%%%%%
%%%%%%%%%%%%%%%%%%%%%%%%%%%%%%%%%%%%%%%%%%%%%%%%%%%%%%%%%%%%%
\section{Correlators in Two-Dimensional CFT: Review}\label{CFTRev}
%%%%%%%%%%%%%%%%%%%%%%%%%%%%%%%%%%%%%%%%%%%%%%%%%%%%%%%%%%%%%
%%%%%%%%%%%%%%%%%%%%%%%%%%%%%%%%%%%%%%%%%%%%%%%%%%%%%%%%%%%%%

One can compare the above analysis of cubic couplings in a higher-spin gravity theory with the structure of three-point correlation functions in a two-dimensional conformal field theory. As explained in \cite{Mkrtchyan:2017ixk,Kessel:2018ugi}, even though the explicit expressions were written down in flat space for simplicity, the classification of cubic vertices is valid for arbitrary Einstein spaces, including AdS space-times. Via the higher-spin AdS/CFT correspondence, the results on cubic couplings then imply that for three higher-spin currents, dual to the corresponding higher-spin gauge fields, we should find two allowed correlation functions, one parity-even and one parity-odd whenever the spins satisfy triangle inequalities~\eqref{TI}. Indeed, the conformal invariance constrains the correlators of higher-spin currents to two structures, but with no sign of a triangle inequality. 

We will review here the well known structure of correlation functions in a conformal field theory (for a textbook see e.g.~\cite{Blumenhagen:2009zz}), and we will resolve the apparent mismatch with the triangle inequalities in the subsequent section by a suitable choice of basis.

All fields in a two-dimensional conformal field theory can be obtained from linear combinations of quasi-primary fields and their derivatives. A quasi-primary field transforms in a simple way under global conformal transformations: if we consider the theory on the complex plane (or better its compactification by including a point at infinity), the global conformal group is the M{\"o}bius group $PSL(2,\mathbb{C})$, and a quasi-primary field $\phi$ transforms under a conformal transformation $z\mapsto w (z)$ as
\begin{equation}
\phi \mapsto \tilde{\phi} \quad ,\quad \tilde{\phi} (w,\bar{w}) = \left(\frac{dw}{dz} \right)^{-h}\left(\frac{d\bar{w}}{d\bar{z}} \right)^{-\bar{h}} \phi (z,\bar{z})\, .
\end{equation}
Here, the real numbers $h$ and $\bar{h}$ are the holomorphic and antiholomorphic conformal weight of the field, respectively. They are related to the scaling dimension $\Delta$ and the spin $s$ by
\begin{equation}
\Delta =h+\bar{h} \quad ,\quad s = |h-\bar{h}|\, .
\end{equation}

Due to the transformation property, the coordinate dependence of a three-point function of three quasi-primary fields is completely fixed,
\begin{multline}
\langle \phi_{1} (z_{1},\bar{z}_{1}) \phi_{2} (z_{2},\bar{z}_{2})\phi_{3} (z_{3},\bar{z}_{3}) \rangle \\
= C\,z_{12}^{h_{3}-h_{1}-h_{2}}\,z_{23}^{h_{1}-h_{2}-h_{3}}\,z_{13}^{h_{2}-h_{1}-h_{3}}\,\bar{z}_{12}^{\bar{h}_{3}-\bar{h}_{1}-\bar{h}_{2}}\,\bar{z}_{23}^{\bar{h}_{1}-\bar{h}_{2}-\bar{h}_{3}}\,\bar{z}_{13}^{\bar{h}_{2}-\bar{h}_{1}-\bar{h}_{3}}\, ,
\end{multline}
where $h_{i},\bar{h}_{i}$ are the conformal weights of the field $\phi_{i}$, and $z_{ij}=z_{i}-z_{j}$. The three-point coefficients $C$ determine the operator product expansion of the fields and thus encode the dynamics of the CFT; all higher-point functions are determined by them.

%The field theory dual to 
A spin-$s$ gauge field in the bulk of asymptotically AdS$_{3}$ space corresponds to a traceless conserved spin-$s$ current $J_{\mu_{1}\dotsb \mu_{s}}$ of scaling dimension $\Delta =s$ in the CFT.  
Due to the tracelessness, only two components of $J$ are non-zero, which in the coordinates $z,\bar{z}$
correspond to $J^{(s)}:=J_{z\dotsb z}$ and $\bar{J}^{(s)}:=J_{\bar{z}\dotsb \bar{z}}$. These are separately conserved, 
\begin{equation}
\partial_{\bar{z}}J^{(s)}=0 \quad ,\quad \partial_{z}\bar{J}^{(s)} =0 \, .
\end{equation}
As for general quasi-primary fields, the coordinate dependence of their three-point correlators is completely fixed --- for three holomorphic fields the form is (see e.g.~\cite[eq.~(2.50)]{Blumenhagen:2009zz})
\begin{equation}\label{3ptholcurrents}
\langle J^{(s_{1})} (z_{1})J^{(s_{2})} (z_{2}) J^{(s_{3})} (z_{3}) \rangle = C_{s_{1},s_{2},s_{3}}\,z_{12}^{s_{3}-s_{1}-s_{2}}\,z_{23}^{s_{1}-s_{2}-s_{3}}\,z_{13}^{s_{2}-s_{1}-s_{3}}\, .
\end{equation}
For every triple of higher-spin currents, there are thus one holomorphic and one antiholomorphic structure for the three-point function from which one can build one parity-even and one parity-odd structure by taking linear combinations. This matches the results for the classification of cubic vertices except that there is no restriction on the possible correlators in form of a triangle inequality: there is no reason that a correlator of three currents with $s_{1}\geq s_{2}+s_{3}$ should vanish.

The mismatch can be resolved by using field redefinitions as we will explain in the following section. The tools that we need there for our explanations are briefly reviewed in the remainder of this section. 

The fields in our theory can be realised as operators acting on a Hilbert space that carries a unitary representation of the conformal group. In a two-dimensional conformal field theory, the conformal algebra can be extended to two copies of the Virasoro algebra with generators $L_{m}$, $\bar{L}_{m}$ satisfying
\begin{equation}\label{Virasoro}
[L_{m},L_{n}] = (m-n)L_{m+n} + \frac{c}{12} m (m^{2}-1)\delta_{m+n,0}
\end{equation}
and the analogous commutation relation for the $\bar{L}_{m}$. $c$ is the central charge.
In a conformal field theory, there is the operator-state correspondence that relates a field $\phi$ to the vector $\phi (0,0)\vac$ where $\vac$ is the $SL (2,\mathbb{C})$ invariant vacuum vector. When we only consider holomorphic fields $J^{(s)}$ the corresponding subspace $\mathcal{H}$ of the Hilbert space carries a representation of the generators $L_{m}$ of one copy of the Virasoro algebras. 

The holomorphic currents $J^{(s)}$ are operators acting on $\mathcal{H}$, and as holomorphic fields they can be expanded as 
\begin{equation}
J^{(s)} ( z) = \sum_{n\in \mathbb{Z}} J^{(s)}_{n}\, z^{-n-s}
\end{equation}
in terms of their modes $J^{(s)}_{n}$. The state corresponding to $J^{(s)}$ is $J^{(s)}_{-s}\vac \in \mathcal{H}$, modes with mode numbers greater than $-s$ annihilate the vacuum,
\begin{equation}
J^{(s)}_{n}\vac   = 0 \quad \text{for}\ n>-s\, .
\end{equation}
Due to the transformation properties of the spin-$s$ current, the commutator of the Virasoro mode $L_{0}$ with any of the modes is
\begin{equation}
[L_{0},J^{(s)}_{m}] = -m\,J_{m}^{(s)} \, .
\end{equation}
The mode $L_{-1}$ corresponds to taking the derivative of the field, and its action on the modes is
\begin{equation}
[L_{-1},J^{(s)}_{m}] = (1-s-m)J^{(s)}_{m-1} \, .
\end{equation}
We assume that we have real fields, $J^{(s)}{}^{\dagger}=J^{(s)}$, such that 
\begin{equation}\label{reality}
J^{(s)}_{n}{}^{\dagger} = J^{(s)}_{-n}\, .
\end{equation}
The coefficient $C_{s_{1},s_{2},s_{3}}$ in the three-point function~\eqref{3ptholcurrents} can then be computed as an expectation value of some of the modes,
\begin{align}
C_{s_{1},s_{2},s_{3}} &= \lim_{z_{1}\to \infty} z_{1}^{2s_{1}} \langle J^{(s_{1})} (z_{1})J^{(s_{2})} (1) J^{(s_{3})} (0) \rangle\\
& = \vacl  J^{(s_{1})}_{s_{1}} \, J^{(s_{2})}_{s_{3}-s_{1}}\,J^{(s_{3})}_{-s_{3}}\vacr \, . 
\end{align}

The currents form the algebraic structure of a $\mathcal{W}$-algebra: the operator product expansion of two holomorphic currents can be expressed in terms of holomorphic currents and their derivatives; conversely, the commutators of the modes of the currents can be expressed again in terms of modes of conserved currents.

The currents dual to the higher-spin gauge fields in the bulk should form a basis of this $\mathcal{W}$-algebra. In the next section, we will show that there is a specific choice for a basis of any $\mathcal{W}$-algebra such that their correlators satisfy the triangle inequalities that was observed for the higher-spin bulk vertices.

%%%%%%%%%%%%%%%%%%%%%%%%%%%%%%%%%%%%%%%%%%%%%%%%%%%%%%%%%%%%%
%%%%%%%%%%%%%%%%%%%%%%%%%%%%%%%%%%%%%%%%%%%%%%%%%%%%%%%%%%%%%
\section{Triangle and Polygon Inequalities for CFT Correlators}\label{sec:CFTtriangle}
%%%%%%%%%%%%%%%%%%%%%%%%%%%%%%%%%%%%%%%%%%%%%%%%%%%%%%%%%%%%%
%%%%%%%%%%%%%%%%%%%%%%%%%%%%%%%%%%%%%%%%%%%%%%%%%%%%%%%%%%%%%

When comparing bulk vertices of higher-spin gauge fields to correlators of currents in the two-dimensional CFT,
we have to be aware that in two dimensions, there is the freedom to redefine the CFT currents by adding products
or derivatives of currents of lower spin. This is in contrast to the situation in higher dimensions. Fixing this ambiguity corresponds to specifying a concrete basis. In this section we will show how to choose a basis for the $\mathcal{W}$-algebra of holomorphic currents in the CFT such that the correlators satisfy triangle inequalities for the spins. We will see that in this basis also the higher-point correlators satisfy analogous polygon inequalities. It is (up to linear transformations) the only basis of quasi-primary fields with this property.

\subsection{The Basis}

We start with a $\mathcal{W}$-algebra generated by a set of elementary fields $\{J^{(s,i)} \mid s\in
\mathcal{S}\subset \mathbb{N},\, 1\leq i\leq m_{s} \}$ of definite spin $s$. The set of all spins that occur is
denoted by $\mathcal{S}$, the number of fields of spin $s$ is denoted by $m_{s}$. The $\mathcal{W}$-algebra consists of all chiral fields $J (z)$ that can be obtained by taking linear combinations of normal ordered products and derivatives of the elementary fields. We assume that the set of elementary fields is minimal in the sense that none of the fields can be obtained from the others by normal-ordered products or derivatives. 

We always assume that the $\mathcal{W}$-algebra contains the holomorphic part $T$ of an energy-momentum tensor (but not necessarily as one of the elementary fields),
\begin{equation}
T (z) = \sum_{m\in \mathbb{Z}}L_{m}\,z^{-m-2}\, ,
\end{equation}
whose modes satisfy the commutation relations of the Virasoro algebra~\eqref{Virasoro}.

The Hilbert space $\mathcal{H}$ generated by all fields is a direct sum of eigenspaces $\mathcal{H}^{(s)}$ with eigenvalue $s\in\mathbb{N}_{0}$ of the $L_{0}$ operator,
\begin{equation}
\mathcal{H} = \bigoplus_{s=0}^{\infty} \mathcal{H}^{(s)}\, .
\end{equation}
The Hilbert space is generated by acting with the modes of the elementary fields on the vacuum vector $\vac$, the subspace $\mathcal{H}^{(s)}$ is then
\begin{equation}
\mathcal{H}^{(s)} = \left\langle J^{(s_{1},i_{1})}_{-n_{1}}\dotsb J^{(s_{r},i_{r})}_{-n_{r}}\vac\ ,\   n_{j}\geq s_{j}, \, \sum_{j=1}^{r}n_{j}=s \right\rangle \, ,
\end{equation}
where the angle brackets denote the linear span. We assume that we have chosen a real basis, see~\eqref{reality}.

Inside $\mathcal{H}^{(s)}$ we now consider the subspace $H^{(s)}\subset \mathcal{H}^{(s)}$ that is generated by all modes of basis fields of spin less than $s$,
\begin{equation}
H^{(s)} = \left\langle J^{(s_{1},i_{1})}_{-n_{1}}\dotsb J^{(s_{r},i_{r})}_{-n_{r}}\vac \ ,\ n_{j}\geq s_{j},\, \sum_{j=1}^{r}n_{j}=s ,\, s_{j}<s \right\rangle \, .
\end{equation}
The co-dimension of $H^{(s)}$ in $\mathcal{H}^{(s)}$ is the number $m_{s}$ of elementary fields of spin~$s$, which by assumption are independent of the products or derivatives of the lower spin fields. We now consider the orthogonal complement $P^{(s)}$ of $H^{(s)}$ in $\mathcal{H}^{(s)}$,
\begin{equation}\label{decomp}
\mathcal{H}^{(s)} = H^{(s)} \oplus P^{(s)} \, ,
\end{equation}
which is of dimension $m_{s}$, and choose a basis $\{\tilde{j}^{(s,1)},\dotsc ,\tilde{j}^{(s,m_{s})} \}$ of $P^{(s)}$. The corresponding fields $\tilde{J}^{(s,i)}$ constitute the basis we are after. 

The space $H^{(s)}$ in terms of fields is the space of all fields that are obtained by taking linear combinations, normal ordered products, and derivatives of fields with spin lower than $s$. It is independent of the original choice of basis fields $J^{(s,i)}$, and the construction of the new basis $\tilde{J}^{(s,i)}$ is therefore unique up to linear transformations within the spaces $P^{(s)}$. 

To illustrate the construction we give a simple example. Consider a CFT with a chiral algebra that is generated by a spin-$1$ current $J$ and the energy-momentum tensor $T$. At spin $1$ there is only one field, so we just take $\tilde{J}^{(1)}=J$. At spin $2$ we have three linearly independent fields,
\begin{equation}
T\quad ,\quad (J\,J)\quad , \quad \partial J \, .
\end{equation}
The subspace $H^{(2)}$ is spanned by the states corresponding to $(J\,J)$ and $\partial J$. In terms of modes it can be written as the span of two vectors,
\begin{equation}
H^{(2)} = \Big \langle J_{-1}\,J_{-1}\Omega\, ,\, J_{-2}\Omega  \Big\rangle\, .
\end{equation}
The commutation relations of the modes are (here we choose a specific normalisation of the field $J$) 
\begin{equation}
[J_{m},J_{n}] = m\,\delta_{m+n,0} \quad ,\quad [L_{m},J_{n}] = -n\,J_{m+n}
\end{equation}
and the standard relations for the Virasoro modes $L_{m}$. It can be easily checked that the orthogonal complement of $H^{(2)}$ in $\mathcal{H}^{(2)}$ is the one-dimensional space
\begin{equation}
P^{(2)} = \left \langle \left(L_{-2} - \frac{1}{2}J_{-1}\,J_{-1}\right)\Omega  \right\rangle\, .
\end{equation}
Therefore we choose the basis field at spin-$2$ to be
\begin{equation}
\tilde{J}^{(2)} = T - \frac{1}{2} (J\,J) \, ,
\end{equation}
i.e.\ the difference of the energy-momentum tensor and the field obtained by the Sugawara construction for $J$. Its modes commute with the modes of $J$,
\begin{equation}
[\tilde{J}^{(2)}_{m},J_{n}] = 0\, ,
\end{equation}
therefore in this basis e.g.\ the three-point function of fields of spin $1$, $1$ and $2$ vanishes, 
\begin{equation}
\langle \tilde{J}^{(2)} \,\tilde{J}^{(1)} \,\tilde{J}^{(1)}\rangle = 0 \, .
\end{equation}
The corresponding statement in the bulk is that there is no cubic coupling of Chern-Simons vector fields\footnote{Note that the chiral currents $\tilde{J}^{(1)}$ here correspond through the $AdS/CFT$
  dictionary to Chern-Simons vector fields in the bulk. Maxwell vector fields, on the other hand, have current couplings to higher spin fields in the bulk. In
  general, interactions of Chern-Simons fields respect the triangle inequalities, while those of Maxwell fields violate them.} to gravity (or any other massless field of spin $s\geq 2$)~\cite{Kessel:2018ugi}.

More sophisticated examples are provided by the $\mathcal{W}_{N}$-algebras which have a basis of fields of spin $2,3,\dotsc ,N$. For spins $3$, $4$ and $5$ the construction just leads to choosing the unique primary field at each spin (we show below that the basis fields $\tilde{J}$ are always primary for spins greater than $2$). At spin $6$ there are two independent primary fields, and just the requirement to choose a primary basis field does not give a unique procedure. There is one primary field at spin $6$ that is constructed out of derivatives and normal-ordered products of fields of lower spins. Our construction is equivalent to choosing a primary field orthogonal to it. Our choice of basis is therefore equivalent to the orthogonality condition that was used e.g.\ in~\cite{Hornfeck:1992tm} in an analysis of $\mathcal{W}$-algebras generated by fields of spin $2,\dotsc ,6$, or in~\cite{Prochazka:2014gqa} in an analysis of the $\mathcal{W}_{\infty}$-algebra.

\subsection{Correlators}

Let us assume that we have chosen a basis as described in the last section, and we denote the basis as $\{J^{(s,i)} \}$ omitting the tilde. We now want to discuss correlators of these fields. Consider the operator product expansion of two fields at the origin evaluated on the vacuum $\vac$,
\begin{equation}
J^{(s_{1},i_{1})} (z) J^{(s_{2},i_{2})} (0)\vac  = \sum_{n\leq s_{2}} z^{-n-s_{1}}\, J^{(s_{1},i_{1})}_{n}\,J^{(s_{2},i_{2})}_{-s_{2}}\vac \, .
\end{equation}
For $n> -s_{1}$ the mode $J^{(s_{1},i_{1})}_{n}$ annihilates the vacuum, whereas for $n\leq -s_{1}$ the mode $J^{(s_{1},i_{1})}_{n}$ generates the $(-s_{1}-n)$th derivative of $J^{(s_{1},i_{1})}$ from the vacuum (up to a factor). When we split the sum, we then obtain
\begin{align}
&J^{(s_{1},i_{1})} (z) J^{(s_{2},i_{2})} (0)\vac \\
&\qquad = \sum_{n=-s_{1}+1}^{s_{2}} z^{-n-s_{1}}\, [J^{(s_{1},i_{1})}_{n},J^{(s_{2},i_{2})}_{-s_{2}}]\vac  + \sum_{n\leq -s_{1}} z^{-n-s_{1}}\, J^{(s_{1},i_{1})}_{n} \,J^{(s_{2},i_{2})}_{-s_{2}}\vac  \\
&\qquad = \sum_{n=1}^{s_{1}+s_{2}} z^{-n}\, [J^{(s_{1},i_{1})}_{-s_{1}+n},J^{(s_{2},i_{2})}_{-s_{2}}] \vac  + \sum_{n\geq 0} \frac{z^{n}}{n!}\big(:\partial^{n} J^{(s_{1},i_{1})}\,J^{(s_{2},i_{2})}:\big)_{-s_{1}-s_{2}-n} \vac  \\
&\qquad\qquad  \in \ \bigoplus_{s=0}^{s_{1}+s_{2}-1} \mathcal{H}^{(s)} \oplus \bigoplus_{s=s_{1}+s_{2}}^{\infty} H^{(s)} \, .
\end{align}
The operator product expansion of two fields $J^{(s_{1},i_{1})}$ and $J^{(s_{2},i_{2})}$ thus only contains fields of spin less than $s_{1}+s_{2}$ and normal ordered products of (derivatives of) $J^{(s_{1},i_{1})}$ and $J^{(s_{2},i_{2})}$.
A field $J^{(s,i)}$ with $s\geq s_{1}+s_{2}$ therefore has by construction vanishing two-point function with any field occurring in the operator product expansion of $J^{(s_{1},i_{1})}$ and $J^{(s_{2},i_{2})}$, and hence the corresponding three-point function vanishes,
\begin{equation}
\langle J^{(s_{1},i_{1})} (z_{1})\, J^{(s_{2},i_{2})} (z_{2})\, J^{(s,i)} (z_{3})\rangle = 0 \quad \text{for}\ s\geq s_{1}+s_{2}\, .
\end{equation}
As the above statement holds for arbitrary spins, the three-point function also vanishes for $s_{1}\geq s+s_{2}$ and $s_{2}\geq s+s_{1}$.
Hence, we get a perfect match with the results for the cubic vertices in the bulk: all three-point functions whose spins violate a triangle inequality vanish. 

A similar argument leads to constraints also for higher correlation functions. We can compute an $r$-point
function by successively taking operator product expansions --- the fields that occur when we take $r-1$ operator products are fields of spin below $s_{1}+\dotsb +s_{r-1}-r+3$, or are normal-ordered products or derivatives of such fields. Hence, the corresponding states are orthogonal to $J^{(s,i)} (0) \vac $ for $s\geq s_{1}+\dotsb +s_{r-1}-r+3$, and the $r$-point correlation function vanishes in this case. The details of the argument leading to this result are presented in Appendix~\ref{app:highercorrelators}.

We conclude that in the specific basis we have chosen the higher CFT correlators also satisfy polygon inequalities: an $r$-point correlation function ($r\geq 3$) of fields of spins $s_{1},\dotsc ,s_{r}$ can be non-zero only if 
\begin{equation}\label{polygonIneq}
s_{k} < s_{1} +\dotsb +s_{k-1} + s_{k+1} \dotsb  + s_{r} -r+3 \quad \text{for all}\ k\in \{1,\dotsc ,r \}\, ,
\end{equation}
or equivalently, if
\begin{equation}
(s_{k}-1) \leq \sum_{\substack{j=1\\ j\not = k}}^{r} (s_{j}-1) \quad \text{for all}\ k\in \{1,\dotsc ,r \}\, .
\end{equation}
This result suggests that we should be able to make even stronger statements for the classification of bulk vertices. Before we study those in Section~\ref{sec:HSpolygon}, we conclude this section by a short discussion of the specific properties of the basis that we constructed.

\subsection{Primary Basis}\label{sec:primary}

It follows from the construction of the new basis that it consists of quasi-primary fields. To see this consider a field $J^{(s,i)}$ and its corresponding state $j^{(s,i)}=J^{(s,i)} (0)\vac $. To show that the field is quasi-primary, we have to show that $L_{1}j^{(s,i)}=0$. This is equivalent to saying that it is orthogonal to all other states of the same $L_{0}$ eigenvalue:
\begin{equation}
\text{for all}\ v\in \mathcal{H}^{(s-1)}:\quad \big\langle v, L_{1}j^{(s,i)}\big\rangle = 0 \, .
\end{equation}
Let $v\in \mathcal{H}^{(s-1)}$. Then
\begin{equation}
\big\langle v, L_{1}j^{(s,i)}\big\rangle = \big\langle L_{-1}v , j^{(s,i)}\big\rangle\, .
\end{equation}
But $L_{-1}v \in H^{(s)}$, because
\begin{equation}
L_{-1} J^{(s_{1},i_{1})}_{-n_{1}} \dotsb J^{(s_{r},i_{r})}_{-n_{r}}\vac  = \sum_{j=1}^{r} (1-s_{j}+n_{j}) J^{(s_{1},i_{1})}_{-n_{1}}\dotsb J^{(s_{j},i_{j})}_{-n_{j}-1}\dotsb J^{(s_{r},i_{r})}_{-n_{r}}\vac  \in H^{(s)}\, ,
\end{equation} 
where all $s_{j}\leq s-1$. Finally, as $j^{(s,i)}$ is orthogonal to $H^{(s)}$, we conclude that
$\big\langle L_{-1}v,j^{(s,i)}\big\rangle=0$, hence $J^{(s,i)}$ is quasi-primary. 

We can perform a similar argument to show that the field is not only quasi-primary, but primary for $s\geq 3$. For that we have to show that $L_{2}j^{(s,i)}=0$, or, equivalently, that $j^{(s,i)}$ is orthogonal to all states in $L_{-2}\mathcal{H}^{(s-2)}$. Indeed, we have
\begin{equation}
L_{-2} \mathcal{H}^{(s-2)}\subset H^{(s)} \quad \text{for}\ s\geq 3 \, .
\end{equation}
This is obvious because vectors in $\mathcal{H}^{(s-2)}$ are constructed from modes of fields of spin less or equal $s-2$, and applying $L_{-2}$ will result in a vector constructed from modes of fields of spin less or equal $s-2$ (and of spin $2$). Thus it is contained in $H^{(s)}$ for $s\geq 3$. 

For the case of spin $2$, we cannot argue like that because $L_{-2}\vac $ generically is not contained in $H^{(2)}$. There are two cases: if $L_{-2}\vac  \in H^{(2)}$, this means that the energy-momentum tensor is obtained from derivatives or normal-ordered products (as e.g.\ in Wess-Zumino-Witten models where the energy momentum tensor is obtained by the Sugawara construction). Then all the elementary spin-$2$ fields $J^{(2,i)}$ in the basis constructed above are orthogonal to $L_{-2}\vac $ and therefore primary. If, on the other hand, $L_{-2}\vac  \not\in H^{(2)}$, then we can choose one of the spin-$2$ fields in the basis as the orthogonal projection of $L_{-2}\vac $ to $P^{(2)}$ (this field is only quasi-primary), and all others can be chosen to be orthogonal to $L_{-2}\vac $ (they are then primary fields). For example, in the CFT dual of Coloured $AdS_3$ Gravity \cite{Gwak:2015vfb}, all of the operators dual to massless spin-$2$ fields will be primaries except for the one dual to the singlet spin-$2$ field -- the metric graviton, whose dual will be quasi-primary -- the stress-energy tensor.
\smallskip

To summarise, we have seen that for any $\mathcal{W}$-algebra, there exists a basis of primary fields (except possibly one quasi-primary spin-2 field) such that the correlators satisfy the polygon inequalities~\eqref{polygonIneq}. It is not difficult to see that the basis is already uniquely characterised by this property: there is only one basis (up to linear transformations) of quasi-primary fields that satisfies the polygon inequalities. This can be shown spin by spin: if at any spin $s$ we choose a field $\tilde{J}^{(s)}$ corresponding to a state that is not orthogonal to all states in $H^{(s)}$, there will be some normal ordered product of (derivatives of) $r$ fields of lower spins (satisfying $s_{1}+\dotsb +s_{r}\leq s$) that has non-vanishing two-point function with $\tilde{J}^{(s)}$. This would imply that the corresponding $r+1$-point function does not vanish --- but for $r\geq 2$ this would violate the polygon inequalities, and for $r=1$ a non-vanishing two-point function with a lower-spin field would be in conflict with the field being quasi-primary.\footnote{Instead of demanding that the basis fields are quasi-primary, one can impose the ``polygon inequalities''~\eqref{polygonIneq} also for $r=2$.}

%%%%%%%%%%%%%%%%%%%%%%%%%%%%%%%%%%%%%%%%%%%%%%%%%%%%%%%%%%%%%
%%%%%%%%%%%%%%%%%%%%%%%%%%%%%%%%%%%%%%%%%%%%%%%%%%%%%%%%%%%%%
\section{Polygon Inequalities for Higher-Order Vertices}\label{sec:HSpolygon}
%%%%%%%%%%%%%%%%%%%%%%%%%%%%%%%%%%%%%%%%%%%%%%%%%%%%%%%%%%%%%
%%%%%%%%%%%%%%%%%%%%%%%%%%%%%%%%%%%%%%%%%%%%%%%%%%%%%%%%%%%%%

Motivated by the CFT results, in particular by the inequalities~\eqref{polygonIneq} for $n$-point functions, we want to see whether in the Noether program we can make also statements about such polygon inequalities for higher-order vertices.
The corresponding inequalities for vertices of order $n$ in the fields would be:
\begin{align}
s_i\leq s_1+s_2+\dots + s_{i-1}+ s_{i+1}+ \dots + s_n-n+2\,, \quad \text{for}\  i=1,\dots , n\,.\label{PI}
\end{align}
For cubic vertices ($n=3$), these inequalities are equivalent to~\eqref{TI}.
It is interesting to note that the above inequalities make sense also for $n=2$. There, $s_1\leq s_2$ and
$s_2\leq s_1$ imply together $s_1=s_2$, which is a statement on the diagonal nature of the quadratic action. 

To show that the polygon inequalities are indeed satisfied under certain assumptions, we proceed by induction and first prove the following statement:\\[2mm]
\textit{Assume that there is a basis of fields, in which the action does not involve any interactions of order
  $k\leq n$ in fields that violate the polygon inequalities 
\begin{align}
s_i \leq s_1+s_2+\dots +s_{i-1}+s_{i+1}+\dots +s_k-k+2\,, \quad \text{for}\  i=1,2,\dots,k\,,\quad k=3,4,\dots,n\label{PIa}
\end{align}
between the spins.
Then, the same is true for interactions involving $n+1$ fields, provided that there are no independent $(n+1)$st order deformations that violate these inequalities.}\\[2mm]
Here, independent deformations of order $n+1$ are solutions to the Noether equation
\begin{align}
\d^0 \mathcal{L}^{(n+1)}\approx 0\,,\label{n+1ID}
\end{align}
where $\approx$ indicates equality up to the free field equations of motion and total derivatives.

The proof of the statement above is simple -- we need to study the Noether equations for gauge invariance:
\begin{align}\label{n+1NE}
\d^0 \mathcal{L}^{(n+1)}+\d^1 \mathcal{L}^{(n)}+\dots+\d^{n-2}\mathcal{L}^{(3)}\approx 0\,.
\end{align}
Assuming that there are no solutions for the homogeneous equation \eqref{n+1ID} that violate the polygon inequalities (assumption of the statement), we can concentrate on the particular solution of~\eqref{n+1NE}. There, any term of $\d^0 \mathcal{L}^{(n+1)}$ should be cancelled by terms of the form $\d^i \mathcal{L}^{(n+1-i)}$. One can write any term of the latter variations in the form
\begin{equation}
\d^i \mathcal{L}^{(n+1-i)} = \sum_{k'} \frac{\d S^{(n+1-i)}}{\d \vf^{k'}} \d^i \vf^{k'} + (\text{total derivative}) \, ,\label{vi}
\end{equation}
where $S^{(n+1-i)}$ is the term in the action that is obtained from $\mathcal{L}^{(n+1-i)}$ by integration.

We have seen in Section~\ref{NFRev} that the triangle inequalities for the cubic coupling induce the same triangle inequalities for the deformation $\delta^{1}$ of the gauge transformation~\eqref{def1}. Similarly, the polygon inequalities for the lower order Lagrangian terms imply analogous inequalities for the gauge transformation deformations $\d^i$. 
When we take these inequalities together, we can deduce that the polygon inequalities are satisfied for the spins entering in any variation of the type \eqref{vi}. 

This can be shown by taking an arbitrary monomial on the right hand side of~\eqref{vi}, where $\frac{\d S^{(n+1-i)}}{\d \vf^{k'}}$ contains fields, say, $\vf^{k_1},\dots \vf^{k_{n-i}}$, and $\d^i \vf^{k'}$ contains fields, say, $\vf^{k_{n-i+1}},\dots,\vf^{k_n}$ and a gauge parameter $\e^{k_{n+1}}$. Then the lower order polygon inequalities~\eqref{PI} for $\mathcal{L}^{(n+1-i)}$ and for $\d^i \vf^{k'}$ imply in particular
\begin{align}
s_{k_1}&\leq s_{k_2}+\dots+s_{k_{n-i}}+s_{k'}-n+i+1\,,\\
s_{k'}&\leq s_{k_{n-i+1}}+\dots+s_{k_{n+1}}-i\,.
\end{align}
Using the second inequality in the first, we obtain
\begin{equation}
s_{k_1}\leq\sum_{i=2}^{n+1} s_{k_i}-n+1 \, ,
\end{equation}
a particular instance of polygon inequalities~\eqref{PI} at order $n+1$. The analogous inequalities for $k_{2},\dotsc$ follow in the same way.
This concludes the proof.

Our statement above can be rephrased as\\[2mm]
{\it Higher-order completions of lower order vertices satisfying the polygon inequalities \eqref{PI} do obey the same polygon inequalities.}\\[-2mm]

The classification of cubic vertices in three dimensions \cite{Mkrtchyan:2017ixk,Kessel:2018ugi} shows that they obey triangle inequalities. If there are no independent higher couplings (or only ones which satisfy polygon inequalities) in a theory, then we can conclude by induction that the following statement is true:\\[2mm]
\textit{For a HS Gravity theory in three dimensions that is fully determined by its cubic Lagrangian, there exists a basis for which all vertices obey polygon inequalities \eqref{PI} for spins $s_i\geq 2$.}\\[-2mm] 

Note that for the proof, we did not need the explicit form of the cubic vertices $V^m_{k,k',k''}$ and the gauge transformations $\d^1_{k',k''}\vf^{(k)}$. We only used the fact that the coefficients
$g^m_{k,k',k''}$ are zero for three spins $s_k,s_{k'},s_{k''}$ that violate the triangle inequalities~\eqref{TI}.

We exclude scalar and Maxwell fields from the discussion, as their interactions do not respect the polygon inequalities already at the cubic order (but the statement also holds for Chern-Simons vector fields). We also excluded higher-order independent deformations that violate the polygon inequalities in the full non-linear theory. This is motivated from the CFT which suggests that no violation of the polygon inequalities occurs. Independent deformations of higher-order will be studied elsewhere.

%%%%%%%%%%%%%%%%%%%%%%%%%%%%%%%%%%%%%%%%%%%%%%%%%%%%%%%%%%%%%
%%%%%%%%%%%%%%%%%%%%%%%%%%%%%%%%%%%%%%%%%%%%%%%%%%%%%%%%%%%%%
\section{Discussion}\label{Discussion}
%%%%%%%%%%%%%%%%%%%%%%%%%%%%%%%%%%%%%%%%%%%%%%%%%%%%%%%%%%%%%
%%%%%%%%%%%%%%%%%%%%%%%%%%%%%%%%%%%%%%%%%%%%%%%%%%%%%%%%%%%%%

We have compared vertices for higher-spin gauge fields in three dimensions with the structure of correlators of conserved currents in a two-dimensional CFT. The classification of cubic vertices is precisely matched by the independent three-point functions in the CFT when one takes into account the possibility of non-linear basis transformations. The CFT analysis has even revealed restrictions on higher correlators: only those for which the spins satisfy polygon inequalities \eqref{PI} can be present. This gives strong predictions for the classification of bulk vertices. 

On the one hand, we expect that higher-order couplings also respect these polygon inequalities. We checked this explicitly for couplings that are continuations of the cubic couplings, but the CFT result implies that the same is true also for independent higher vertices. We will study this problem elsewhere.

On the other hand, the CFT results are full quantum results, so one expects that in the bulk theory --- as long as higher-spin gauge invariance is preserved --- the polygon inequalities are even satisfied if we take quantum corrections into account. Indeed it is straightforward to show (by an argument similar to the one used in Section~\ref{sec:HSpolygon}) that in a theory of fields of spin greater or equal to $2$, any tree-level Witten diagram built from vertices that satisfy polygon inequalities will vanish if the spins of the external fields do not satisfy the polygon inequalities \eqref{PI}. Similar conclusions cannot be made for loop diagrams though. 

Whereas from the pure CFT perspective all different bases appear on equal footing,
from the holographic perspective the basis we identified seems physically the most sensible choice: it directly reproduces the structure of cubic on-shell vertices. It is interesting to identify the significance of this choice directly in a holographic setting. When we consider as a simple example a Chern-Simons formulation of higher-spin gravity with a gauge algebra $\mathfrak{g}\oplus \mathfrak{g}$, its asymptotic symmetry algebra is a classical $\mathcal{W}$-algebra that is the Drinfeld-Sokolov reduction \cite{Drinfeld:1984qv} of the affine Lie algebra $\widehat{\mathfrak{g}}_{k}\oplus \widehat{\mathfrak{g}}_{k}$ (where the level depends on the level of the Chern-Simons theory). The Drinfeld-Sokolov reduction can be performed in different gauges, the gauge fixing determines the basis in which the $\mathcal{W}$-algebra is realised. Performing the reduction in the highest-weight gauge,\footnote{There are two other common gauge choices for $\mathfrak{g}=\mathfrak{sl}(N)$: the \textsl{u-gauge} which leads to a particularly simple realisation of the $\mathcal{W}$-algebra where the highest non-linearities are quadratic~\cite{Dickey:book}, and the \textsl{diagonal gauge} that leads to a free field realisation~\cite{Balog:1990mu} which therefore can be a good starting point to discuss quantisation~\cite{Campoleoni:2017xyl}.} one arrives at a primary basis (except for a quasi-primary spin-$2$ field). Furthermore the Poisson brackets (that in the quantum $\mathcal{W}$-algebra turn into commutators which encode the singular operator products of the fields) in this basis are constrained by the representation theory of the embedded $\mathfrak{sl}(2)$ algebra (which corresponds to the gravity part). The Poisson bracket takes the form~\cite{Campoleoni:2011hg}
\begin{equation}
\{J^{(s_{1})},J^{(s_{2})} \} = c_{s_{1}}\,\delta_{s_{1},s_{2}} + \left(\begin{array}{l}
\text{fields of spin}\ s\ \text{with}\\
|s_{1}-s_{2}|<s<s_{1}+s_{2}
\end{array}\right) \, .
\end{equation}
The bound from above on the possible spins appearing in the bracket is satisfied in any basis, but the bound from below is special. If it is still present in the quantum $\mathcal{W}$-algebra, the three-point correlators will satisfy triangle inequalities. The highest-weight gauge in the Drinfeld-Sokolov reduction thus leads to the classical version of the primary basis that we considered in this paper; it seems to be a preferred gauge choice to discuss holography.

\section*{Acknowledgements}

We thank Pan Kessel for useful discussions. S.F.\ thanks the Max-Planck-Institute for Gravitational Physics in Potsdam for hospitality during the workshop on `Higher Symmetries and Quantum Gravity' which gave the opportunity to discuss and advance the work described in this article.

\appendix

\section{Higher Correlators}\label{app:highercorrelators}

In this appendix, we provide the argument that for $r\geq 2$
\begin{equation}
J^{(s_{1},i_{1})} (z_{1}) \dotsb J^{(s_{r},i_{r})} (z_{r})\vac \,\in \bigoplus_{s=0}^{S-r+1} \mathcal{H}^{(s)} \oplus \bigoplus_{s=S-r+2}^{\infty} H^{(s)}\, ,
\end{equation}
where $S=s_{1}+\dotsb +s_{r}$. This can be equivalently stated in terms of modes as
\begin{equation}\label{higheroperatorproducts}
J^{(s_{1},i_{1})}_{-s_{1}+n_{1}} \dotsb J^{(s_{r},i_{r})}_{-s_{r}+n_{r}}\vac  \,\in \bigoplus_{s=0}^{S-r+1} \mathcal{H}^{(s)} \oplus \bigoplus_{s=S-r+2}^{\infty} H^{(s)}
\end{equation}
for all $n_{k}\in\mathbb{Z}$. The idea to prove this is the following: if all $n_{k}$ are negative or zero, we immediately see that the vector is in the space generated by normal ordered products of derivatives of the fields $J^{(s_{k},i_{k})}$; hence it is a state of $L_{0}$-weight $S-\sum n_{k}\geq S$ that is generated by fields of spin less than $\max (s_{1},\dotsc ,s_{r})<S$. If one or more of the $n_{k}$ are positive, we commute them to the right until they annihilate the vacuum vector $\vac$. Using general structures for the commutation relations (reviewed below), one can see that the spins of the fields that are generated in this process are bounded such that no fields of spin $S-r+2$ or higher are produced.

We have seen in Section~\ref{sec:primary} that the fields $J^{(s,i)}$ in our basis are quasi-primary. Commutation relations of general holomorphic quasi-primary fields always take the form (see e.g.~\cite[eq.~(2.54)]{Blumenhagen:2009zz})
\begin{equation}\label{commrel}
[\phi_{(i)m},\phi_{(j)n}] = \sum_{\substack{k\\ s_{k}<s_{i}+s_{j}}} C_{ij}^{k}\,p_{ijk} (m,n)\,\phi_{(k)m+n} + d_{ij}\,\delta_{m,-n} \binom{m+s_{i}-1}{2s_{i}-1}\,,
\end{equation}
where $\phi_{(i)m}$ denotes the $m^{\text{th}}$ mode of a field $\phi_{(i)}$ of spin $s_{i}$ and $C_{ij}^{k}$ are structure constants. The $p_{ijk} (m,n)$ are polynomials in $m$ and $n$ which are completely fixed by the spins $s_{i}$, $s_{j}$ and $s_{k}$. In our case, the basis fields $J^{(s,i)}$ constitute some of the $\phi_{(j)}$, the remaining ones are obtained by quasi-primary combinations of normal ordered products and derivatives of the basis fields. The modes $\phi_{(k)m}$ of such a non-elementary field can be expressed as sums of products of modes of the basis fields, where the sum of the spins of the constituent fields is always lower or equal to the spin $s_{k}$ of the field.

Whenever in an expression like~\eqref{higheroperatorproducts} we commute one mode $J^{(s_{1},i_{1})}$ with another one $J^{(s_{2},i_{2})}$, the quantity `sum of spins of generators $-$ number of generators' is never increased due to the commutation relations~\eqref{commrel}. As on the other hand the `sum of spins of generators' is decreased in the commutator~\eqref{commrel}, the process of commuting generators to the right has to stop at some point. The spin $s$ of a single generator that could remain therefore satisfies 
\begin{equation}
s-1\leq S-r \qquad \Longrightarrow \qquad s\leq S-r+1\,,
\end{equation} 
which completes the proof.

%\bibliographystyle{unsrt}
%\bibliography{papers}

\begin{thebibliography}{999}

\bibitem{Vasiliev:1990en}
M.~A. Vasiliev, 
{\it ``Consistent equation for interacting gauge fields of all spins in (3+1)-dimensions,''}  
\href{http://www.sciencedirect.com/science/article/pii/0370269390914006}{{\it Phys. \ Lett. \ B} {\bf B243} (1990) 378}.
  %%CITATION = doi:10.1016/0370-2693(90)91400-6;%%

\bibitem{Prokushkin:1998bq}
  S.~F.~Prokushkin and M.~A.~Vasiliev,
  {\it ``Higher spin gauge interactions for massive matter fields in 3-D AdS space-time,''}
\href{https://doi.org/10.1016/S0550-3213(98)00839-6}{{\it Nucl.\ Phys.\ B} {\bf 545} (1999) 385};
\href{http://arxiv.org/abs/hep-th/9806236}{\tt arXiv:hep-th/9806236}.
  %%CITATION = doi:10.1016/S0550-3213(98)00839-6;%%

\bibitem{Vasiliev:2003ev}
  M.~A.~Vasiliev,
 {\it ``Nonlinear equations for symmetric massless higher spin fields in (A)dS(d)''},
\href{http://www.sciencedirect.com/science/article/pii/S0370269303008724}{{\it Phys.\ Lett.\ B} {\bf 567} (2003) 139}; 
\href{http://xxx.lanl.gov/abs/hep-th/0304049}{{\tt  arXiv:hep-th/0304049}}.

\bibitem{bciv}
  X.~Bekaert, S.~Cnockaert, C.~Iazeolla and M.~A.~Vasiliev,
{\it ``Nonlinear higher spin theories in various dimensions},''
\href{http://xxx.lanl.gov/abs/hep-th/0503128}{{\tt  hep-th/0503128}} .
  %%CITATION = HEP-TH/0503128;%%

 %\cite{Campoleoni:2011tn}
\bibitem{Campoleoni:2011tn}
  A.~Campoleoni,
 {\it ``Higher Spins in D = 2 + 1,''}
 \href{http://www.worldscientific.com/doi/abs/10.1142/9789814522519_0020}{{\it Subnucl.\ Ser.}\  {\bf 49} (2013) 385};
%  doi:10.1142/9789814522519_0020
\href{https://arxiv.org/abs/1110.5841}{\tt  arXiv:1110.5841}.
  %%CITATION = doi:10.1142/9789814522519_0020;%%
  %19 citations counted in INSPIRE as of 28 Dec 2017
  
\bibitem{Didenko:2014dwa}
  V.~E.~Didenko and E.~D.~Skvortsov,
  {\it ``Elements of Vasiliev theory,''}
\href{https://arxiv.org/abs/1401.2975}{\tt  arXiv:1401.2975}.
  %%CITATION = ARXIV:1401.2975;%%

\bibitem{Rahman:2015pzl}
  R.~Rahman and M.~Taronna,
  {\it ``From Higher Spins to Strings: A Primer,''}
  \href{http://arxiv.org/abs/arXiv:1512.07932}{arXiv:1512.07932}.
  %%CITATION = ARXIV:1512.07932;%%
 
 %\cite{Kessel:2017mxa}
\bibitem{Kessel:2017mxa}
  P.~Kessel,
  {\it ``The Very Basics of Higher-Spin Theory,''}
  \href{https://pos.sissa.it/296/001}{{\it PoS Modave} {\bf 2016} (2017) 001};
%  doi:10.22323/1.296.0001
\href{http://arxiv.org/abs/arXiv:1702.03694}{arXiv:1702.03694}.
  %%CITATION = doi:10.22323/1.296.0001;%%
  %1 citations counted in INSPIRE as of 24 Dec 2018

\bibitem{Giombi:2016ejx}
  S.~Giombi,
  {\it ``Higher Spin — CFT Duality,''}
  \href{https://doi.org/10.1142/9789813149441_0003}{{\it New Frontiers in Fields and Strings} (2017) 137};
  \href{http://arxiv.org/abs/arXiv:1607.02967}{\tt arXiv:1607.02967}.
  %%CITATION = doi:10.1142/9789813149441_0003;%%

%\cite{Joung:2014qya}
\bibitem{Joung:2014qya}
  E.~Joung and K.~Mkrtchyan,
 {\it ``Notes on higher-spin algebras: minimal representations and structure constants,''}
\href{https://link.springer.com/article/10.1007\%2FJHEP05\%282014\%29103}{{\it JHEP} {\bf 1405} (2014) 103};
\href{https://arxiv.org/abs/1401.7977}{\tt  arXiv:1401.7977}.
  %%CITATION = doi:10.1007/JHEP05(2014)103;%%
  %20 citations counted in INSPIRE as of 26 Feb 2017
       
\bibitem{Henneaux:2010xg}
  M.~Henneaux and S.~J.~Rey,
  {\it ``Nonlinear $W_{infinity}$ as Asymptotic Symmetry of Three-Dimensional Higher Spin Anti-de Sitter Gravity,''}
\href{https://doi.org/10.1007/JHEP12(2010)007}{{\it JHEP} {\bf 1012} (2010) 007};
 \href{http://arxiv.org/abs/arXiv:1008.4579}{\tt arXiv:1008.4579}.
  %%CITATION = doi:10.1007/JHEP12(2010)007;%%

%\cite{Campoleoni:2010zq}
\bibitem{Campoleoni:2010zq}
  A.~Campoleoni, S.~Fredenhagen, S.~Pfenninger and S.~Theisen,
 {\it ``Asymptotic symmetries of three-dimensional gravity coupled to higher-spin fields,''}
\href{https://link.springer.com/article/10.1007%2FJHEP11%282010%29007}{{\it JHEP} {\bf 1011} (2010) 007};
%  doi:10.1007/JHEP11(2010)007
\href{https://arxiv.org/abs/1008.4744}{\tt  arXiv:1008.4744}.
  %%CITATION = doi:10.1007/JHEP11(2010)007;%%
  %330 citations counted in INSPIRE as of 27 Dec 2017       
 
\bibitem{Gaberdiel:2011wb}
  M.~R.~Gaberdiel and T.~Hartman,
  {\it ``Symmetries of Holographic Minimal Models,''}
\href{https://doi.org/10.1007/JHEP05(2011)031}{{\it JHEP} {\bf 1105} (2011) 031};
\href{http://arxiv.org/abs/arXiv:1101.2910}{\tt arXiv:1101.2910}.
  %%CITATION = doi:10.1007/JHEP05(2011)031;%%

\bibitem{Campoleoni:2011hg}
  A.~Campoleoni, S.~Fredenhagen and S.~Pfenninger,
  {\it ``Asymptotic W-symmetries in three-dimensional higher-spin gauge theories,''}
\href{https://doi.org/10.1007/JHEP09(2011)113}{{\it JHEP} {\bf 1109} (2011) 113};
\href{http://arxiv.org/abs/arXiv:1107.0290}{\tt arXiv:1107.0290}.
  %%CITATION = doi:10.1007/JHEP09(2011)113;%%

\bibitem{Campoleoni:2014tfa}
  A.~Campoleoni and M.~Henneaux,
{\it ``Asymptotic symmetries of three-dimensional higher-spin gravity: the metric approach,''}
\href{http://link.springer.com/article/10.1007%2FJHEP03%282015%29143}{{\it JHEP} {\bf 1503} (2015) 143};
\href{https://arxiv.org/abs/1412.6774}{\tt  arXiv:1412.6774}.
  %%CITATION = doi:10.1007/JHEP03(2015)143;%%  
  
%\cite{Joung:2017hsi}
\bibitem{Joung:2017hsi}
  E.~Joung, J.~Kim, J.~Kim and S.~J.~Rey,
 {\it ``Asymptotic Symmetries of Colored Gravity in Three Dimensions,''}
\href{https://link.springer.com/article/10.1007%2FJHEP03%282018%29104}{{\it JHEP} {\bf 1803} (2018) 104}; 
 \href{https://arxiv.org/abs/1712.07744}{\tt  arXiv:1712.07744}.
  %%CITATION = ARXIV:1712.07744;%%

%\cite{Gaberdiel:2010pz}
\bibitem{Gaberdiel:2010pz}
  M.~R.~Gaberdiel and R.~Gopakumar,
{\it  ``An $AdS_3$ Dual for Minimal Model CFTs,''}
\href{https://journals.aps.org/prd/abstract/10.1103/PhysRevD.83.066007}{{\it Phys.\ Rev.\ D} {\bf 83} (2011) 066007};
%  doi:10.1103/PhysRevD.83.066007
\href{https://arxiv.org/abs/1011.2986}{\tt  arXiv:1011.2986}. 
  %%CITATION = doi:10.1103/PhysRevD.83.066007;%%
  %313 citations counted in INSPIRE as of 27 Dec 2017       

%\cite{Gaberdiel:2012uj}
\bibitem{Gaberdiel:2012uj}
  M.~R.~Gaberdiel and R.~Gopakumar,
{\it ``Minimal Model Holography,''}
\href{http://iopscience.iop.org/article/10.1088/1751-8113/46/21/214002/meta}{{\it J.\ Phys.\ A} {\bf 46} (2013) 214002};
%  doi:10.1088/1751-8113/46/21/214002
  \href{https://arxiv.org/abs/1207.6697}{\tt  arXiv:1207.6697}. 
  %%CITATION = doi:10.1088/1751-8113/46/21/214002;%%
  %202 citations counted in INSPIRE as of 27 Dec 2017

%\cite{Fronsdal:1978rb}
\bibitem{Fronsdal:1978rb}
  C.~Fronsdal,
  {\it ``Massless Fields with Integer Spin,''}
\href{https://journals.aps.org/prd/abstract/10.1103/PhysRevD.18.3624}{ {\it  Phys.\ Rev.\ D} {\bf 18} (1978) 3624.}
  %doi:10.1103/PhysRevD.18.3624
  %%CITATION = doi:10.1103/PhysRevD.18.3624;%%
  %612 citations counted in INSPIRE as of 23 Dec 2017
  
\bibitem{Mkrtchyan:2017ixk}
  K.~Mkrtchyan,
  {\it ``Cubic interactions of massless bosonic fields in three dimensions,''}
  \href{https://doi.org/10.1103/PhysRevLett.120.221601}{{\it Phys.\ Rev.\ Lett.\ } {\bf 120} (2018) no.22,  221601};
  \href{https://arxiv.org/abs/1712.10003}{\tt arXiv:1712.10003}.
  %%CITATION = doi:10.1103/PhysRevLett.120.221601;%%  
  
\bibitem{Kessel:2018ugi}
  P.~Kessel and K.~Mkrtchyan,
  {\it ``Cubic interactions of massless bosonic fields in three dimensions II: Parity-odd and Chern-Simons vertices,''}
  \href{https://doi.org/10.1103/PhysRevD.97.106021}{{\it Phys.\ Rev.\ D} {\bf 97} (2018) no.10,  106021};
  \href{https://arxiv.org/abs/1803.02737}{\tt arXiv:1803.02737}.
  %%CITATION = doi:10.1103/PhysRevD.97.106021;%%  
  
%\cite{Blencowe:1988gj}
\bibitem{Blencowe:1988gj}
  M.~P.~Blencowe,
{\it ``A Consistent Interacting Massless Higher Spin Field Theory in $D$ = (2+1),''}
\href{http://iopscience.iop.org/article/10.1088/0264-9381/6/4/005/meta}{{\it Class.\ Quant.\ Grav.}\  {\bf 6} (1989) 443.}
%  doi:10.1088/0264-9381/6/4/005
  %%CITATION = doi:10.1088/0264-9381/6/4/005;%%
  %212 citations counted in INSPIRE as of 27 Dec 2017 

%\cite{Gwak:2015vfb}
\bibitem{Gwak:2015vfb}
  S.~Gwak, E.~Joung, K.~Mkrtchyan and S.~J.~Rey,
{\it ``Rainbow Valley of Colored (Anti) de Sitter Gravity in Three Dimensions,''}
  \href{http://link.springer.com/article/10.1007\%2FJHEP04\%282016\%29055}{\it JHEP {\bf 1604} (2016) 055};
  %doi:10.1007/JHEP04(2016)055
  \href{https://arxiv.org/abs/1511.05220}{\tt  arXiv:1511.05220}.
  %%CITATION = doi:10.1007/JHEP04(2016)055;%%
  %8 citations counted in INSPIRE as of 31 Oct 2016

%\cite{Gwak:2015jdo}
\bibitem{Gwak:2015jdo}
  S.~Gwak, E.~Joung, K.~Mkrtchyan and S.~J.~Rey,
{\it ``Rainbow vacua of colored higher-spin (A)dS$_{3}$ gravity,''}
  \href{http://link.springer.com/article/10.1007\%2FJHEP05\%282016\%29150}{\it JHEP {\bf 1605} (2016) 150};
  %doi:10.1007/JHEP05(2016)150
  \href{https://arxiv.org/abs/1511.05975}{\tt  arXiv:1511.05975}.
  %%CITATION = doi:10.1007/JHEP05(2016)150;%%
  %8 citations counted in INSPIRE as of 31 Oct 2016

%\cite{Campoleoni:2012hp}
\bibitem{Campoleoni:2012hp}
  A.~Campoleoni, S.~Fredenhagen, S.~Pfenninger and S.~Theisen,
 {\it ``Towards metric-like higher-spin gauge theories in three dimensions,''}
  \href{http://iopscience.iop.org/article/10.1088/1751-8113/46/21/214017/meta}{{\it J.\ Phys.\ A} {\bf 46} (2013) 214017};
  %doi:10.1088/1751-8113/46/21/214017
  \href{https://arxiv.org/abs/1208.1851}{\tt  arXiv:1208.1851}.
  %%CITATION = doi:10.1088/1751-8113/46/21/214017;%%
  %67 citations counted in INSPIRE as of 27 Dec 2017

%\cite{Fredenhagen:2014oua}
\bibitem{Fredenhagen:2014oua}
  S.~Fredenhagen and P.~Kessel,
 {\it ``Metric- and frame-like higher-spin gauge theories in three dimensions,''}
\href{http://iopscience.iop.org/article/10.1088/1751-8113/48/3/035402/meta}{{\it J.\ Phys.\ A} {\bf 48} (2015) no.3,  035402};
 % doi:10.1088/1751-8113/48/3/035402
 \href{https://arxiv.org/abs/1408.2712}{\tt  arXiv:1408.2712}.
  %%CITATION = doi:10.1088/1751-8113/48/3/035402;%%
  %13 citations counted in INSPIRE as of 27 Dec 2017

\bibitem{Metsaev:2005ar}
  R.~R.~Metsaev,
{\it ``Cubic interaction vertices of massive and massless higher spin fields,''}
\href{http://www.sciencedirect.com/science/article/pii/S0550321306007978}{ {\it Nucl.\ Phys.\ B} {\bf 759} (2006) 147}; 
\href{https://arxiv.org/abs/hep-th/0512342}{\tt  hep-th/0512342}.
  %%CITATION = doi:10.1016/j.nuclphysb.2006.10.002;%%

\bibitem{Manvelyan:2010jr}
  R.~Manvelyan, K.~Mkrtchyan and W.~R\"uhl,
{\it ``General trilinear interaction for arbitrary even higher spin gauge fields,''}
\href{http://www.sciencedirect.com/science/article/pii/S0550321310002427}{ {\it Nucl.\ Phys.\ B} {\bf 836} (2010) 204};
\href{https://arxiv.org/abs/1003.2877}{\tt  arXiv:1003.2877}.
  %%CITATION = doi:10.1016/j.nuclphysb.2010.04.019;%%

\bibitem{Giombi:2011rz}
  S.~Giombi, S.~Prakash and X.~Yin,
  {\it ``A Note on CFT Correlators in Three Dimensions,''}
\href{https://doi.org/10.1007/JHEP07(2013)105}{{\it JHEP} {\bf 1307} (2013) 105};
\href{http://arxiv.org/abs/arXiv:1104.4317}{\tt arXiv:1104.4317}.
  %%CITATION = doi:10.1007/JHEP07(2013)105;%%

\bibitem{Costa:2011mg}
  M.~S.~Costa, J.~Penedones, D.~Poland and S.~Rychkov,
  {\it ``Spinning Conformal Correlators,''}
  \href{https://doi.org/10.1007/JHEP11(2011)071}{{\it JHEP} {\bf 1111} (2011) 071};
  \href{http://arxiv.org/abs/arXiv:1107.3554}{\tt arXiv:1107.3554}.
  %%CITATION = doi:10.1007/JHEP11(2011)071;%%

\bibitem{Joung:2011ww}
  E.~Joung and M.~Taronna,
{\it ``Cubic interactions of massless higher spins in (A)dS: metric-like approach,''}
\href{http://www.sciencedirect.com/science/article/pii/S0550321312001642}{\color{bleuf} {\it Nucl.\ Phys.\ B} {\bf 861} (2012) 145};
\href{https://arxiv.org/abs/1110.5918}{\tt \color{bleuf} arXiv:1110.5918}.
  %%CITATION = doi:10.1016/j.nuclphysb.2012.03.013;%%
 
\bibitem{Conde:2016izb}
  E.~Conde, E.~Joung and K.~Mkrtchyan,
{\it  ``Spinor-Helicity Three-Point Amplitudes from Local Cubic Interactions,''}
\href{http://link.springer.com/article/10.1007%2FJHEP08%282016%29040}{{\it JHEP} {\bf 1608} (2016) 040};
\href{https://arxiv.org/abs/1605.07402}{\tt  arXiv:1605.07402}.
  %%CITATION = doi:10.1007/JHEP08(2016)040;%%
  
%\cite{Francia:2016weg}
\bibitem{Francia:2016weg}
  D.~Francia, G.~Lo Monaco and K.~Mkrtchyan,
{\it ``Cubic interactions of Maxwell-like higher spins,''}
  \href{https://link.springer.com/article/10.1007%2FJHEP04%282017%29068}{{\it JHEP} {\bf 1704} (2017) 068};
%  doi:10.1007/JHEP04(2017)068
\href{https://arxiv.org/abs/1611.00292}{\tt  arXiv:1611.00292}.
  %%CITATION = doi:10.1007/JHEP04(2017)068;%%
  %7 citations counted in INSPIRE as of 27 Dec 2017
 
\bibitem{Kravchuk:2016qvl}
  P.~Kravchuk and D.~Simmons-Duffin,
 {\it ``Counting Conformal Correlators,''}
\href{https://doi.org/10.1007/JHEP02(2018)096}{{\it JHEP} {\bf 1802} (2018) 096};
\href{http://arxiv.org/abs/arXiv:1612.08987}{\tt arXiv:1612.08987}.
  %%CITATION = doi:10.1007/JHEP02(2018)096;%%

 %\cite{Manvelyan:2010wp}
\bibitem{Manvelyan:2010wp}
  R.~Manvelyan, K.~Mkrtchyan and W.~Ruehl,
{\it ``Direct Construction of A Cubic Selfinteraction for Higher Spin gauge Fields,''}
\href{http://www.sciencedirect.com/science/article/pii/S0550321310005961?via%3Dihub}{{\it Nucl.\ Phys.\ B} {\bf 844} (2011) 348};
%  doi:10.1016/j.nuclphysb.2010.11.009
\href{https://arxiv.org/abs/1002.1358}{\tt  arXiv:1002.1358}.
  %%CITATION = doi:10.1016/j.nuclphysb.2010.11.009;%%
  %33 citations counted in INSPIRE as of 28 Dec 2017    
  
%\cite{Manvelyan:2009vy}
\bibitem{Manvelyan:2009vy}
  R.~Manvelyan, K.~Mkrtchyan and W.~Ruhl,
{\it ``Off-shell construction of some trilinear higher spin gauge field interactions,''}
\href{http://www.sciencedirect.com/science/article/pii/S0550321309003629?via%3Dihub}{{\it Nucl.\ Phys.\ B} {\bf 826} (2010) 1};
 % doi:10.1016/j.nuclphysb.2009.07.007
\href{http://xxx.lanl.gov/abs/0903.0243}{{\tt  arXiv:0903.0243}}.
  %%CITATION = doi:10.1016/j.nuclphysb.2009.07.007;%%
  %26 citations counted in INSPIRE as of 28 Dec 2017

\bibitem{Blumenhagen:2009zz}
  R.~Blumenhagen and E.~Plauschinn,
  {\it ``Introduction to conformal field theory : with applications to String theory,''}
  \href{https://doi.org/10.1007/978-3-642-00450-6}{{\it Lect.\ Notes Phys.\ }{\bf 779} (2009) 1}.
  %%CITATION = doi:10.1007/978-3-642-00450-6;%%



\bibitem{Hornfeck:1992tm}
  K.~Hornfeck,
  {\it ``W algebras with set of primary fields of dimensions (3, 4, 5) and (3, 4, 5, 6),''}
\href{https://doi.org/10.1016/0550-3213(93)90281-S}{{\it Nucl.\ Phys.\ B} {\bf 407} (1993) 237};
 \href{http://arxiv.org/abs/hep-th/9212104}{\tt arXiv:hep-th/9212104}.
  %%CITATION = doi:10.1016/0550-3213(93)90281-S;%%

\bibitem{Prochazka:2014gqa}
  T.~Procházka,
  {\it ``Exploring $ {\mathcal{W}}_{\infty } $ in the quadratic basis,''}
  \href{https://doi.org/10.1007/JHEP09(2015)116}{{\it JHEP} {\bf 1509} (2015) 116};
  \href{https://arxiv.org/abs/1411.7697}{\tt arXiv:1411.7697}.
  %%CITATION = doi:10.1007/JHEP09(2015)116;%%
  
  \bibitem{Drinfeld:1984qv}
  V.~G.~Drinfeld and V.~V.~Sokolov,
  {\it ``Lie algebras and equations of Korteweg-de Vries type,''}
  \href{https://doi.org/10.1007/BF02105860}{{\it J.\ Sov.\ Math.}\  {\bf 30} (1984) 1975}.
  %doi:10.1007/BF02105860
  %%CITATION = doi:10.1007/BF02105860;%%

\bibitem{Dickey:book}
L.A.~Dickey, {\it ``Soliton equations and Hamiltonian systems,''}
{\it World Scientific}, Singapore (2003)

\bibitem{Balog:1990mu}
  J.~Balog, L.~Feher, L.~O'Raifeartaigh, P.~Forgacs and A.~Wipf,
  {\it ``Toda Theory and $W$ Algebra From a Gauged {WZNW} Point of View,''}
  \href{https://doi.org/10.1016/0003-4916(90)90029-N}{{\it Annals Phys.}\  {\bf 203} (1990) 76}.
  %%CITATION = doi:10.1016/0003-4916(90)90029-N;%%

%\cite{Campoleoni:2017xyl}
\bibitem{Campoleoni:2017xyl}
  A.~Campoleoni, S.~Fredenhagen and J.~Raeymaekers,
{\it ``Quantizing higher-spin gravity in free-field variables,''}
\href{https://link.springer.com/article/10.1007%2FJHEP02%282018%29126}{{\it JHEP}\  {\bf 1802} (2018) 126};
  \href{https://arxiv.org/abs/1712.08078}{\tt  arXiv:1712.08078}.
  %%CITATION = ARXIV:1712.08078;%%  




\end{thebibliography}

\end{document}